\documentclass[aps,prd,showpacs]{revtex4}
\usepackage{graphicx}
\usepackage{psfig}
\usepackage{revsymb}
\begin{document}
\title{ Thomas-Fermi Model for a Bulk  Self-Gravitating Stellar Object in Two-Dimension}
\author{Sanchari De  and Somenath Chakrabarty$^\dagger$}
\affiliation{Department of Physics, Visva-Bharati, Santiniketan-731235, 
India\\
$^\dagger$somenath.chakrabarty@visva-bharati.ac.in\\}
\begin{abstract}
In this article we have solved an hypothetical problem related to
the stability and gross properties of two dimensional self-gravitating 
stellar objects using Thomas-Fermi model. The formalism presented here is 
an extension of the standard three-dimensional problem discussed in the 
book on statistical physics, Part-I  by Landau and Lifshitz. Further, the formalism presented in this article may be
considered as class problem for post-graduate level students of
physics or may be assigned as a part of their dissertation project.
\end{abstract}
\maketitle
\section{Introduction}
The study of gross properties of bulk self-gravitating objects using 
Thomas-Fermi 
model has been discussed in a very lucid manner
in the text book on Statistical Physics by Landau and Lifshitz \cite{LL}. 
In this book the model has also been extended for the
bulk system with ultra-relativistic electrons as one of the constituents. 
Analogous to the conventional white dwarf
model, these electrons in both non-relativistic and
ultra-relativistic cases provide degeneracy pressure to make the bulk system 
stable against gravitational collapse. The results are therefore an
alternative to Lane-Emden equation or Chandrasekhar equation
\cite{ST,RQ}. The 
mathematical formalism along with the numerical estimates of various 
parameters, e.g., mass, radius, etc. for white
dwarf stars may be taught as standard astrophysical problems in the
master of science level classes for the students having astrophysics
and cosmology as
spacial paper. The standard version of Thomas-Fermi model has also been 
taught in the M.Sc level atomic physics and 
statistical mechanics general classes.

More than two decades ago, Bhaduri et.al. \cite{AJP} have developed a
formalism for Thomas-Fermi model
for a two dimensional atom. The problem can also be given to the
advanced level M.Sc. students
in the quantum mechanics classes. The numerical evaluation of various
quantities
associated with the two dimensional atoms are also found to be useful 
for the`
students to learn numerical techniques, computer programing along with the
physics  of the problem. The work of Bhaduri et.al. is an extension of
standard Thomas-Fermi model for heavy atoms into two-dimensional
scenario. A two-dimensional version of Thomas-Fermi model has
also been used to study the stability and some of the gross properties of two-dimensional star cluster \cite{MAX}.
In our opinion this is the first attempt to apply Thomas-Fermi model to two-dimensional gravitating object. However, to the best of our 
knowledge the two-dimensional generalization of Thomas-Fermi model to study
gross properties of bulk self-gravitating objects, e.g., white dwarfs has not been reported earlier. This problem can also be 
treated as
standard M.Sc level class problem for the advanced level students with 
Astrophysics and Cosmology as special paper.
In this article we shall therefore develop a formalism for two dimensional 
version of Thomas-Fermi model to investigate some of the gross properties 
of two-dimensional hypothetical white dwarf stars. The work is essentially 
an extension of the standard three-dimensional problem which is discussed in
the statistical physics book by Landau and Lifshitz \cite{LL}. The motivation of this work is to study Newtonian
gravity in two-dimension. Analogous Coulomb problem with logarithmic type potential has been investigated in an
extensive manner. However, the identical problem for gravitating objects has not been thoroughly studied (except in
\cite{MAX}). One can use this two-dimensional gravitational picture
as a model calculation to study the stability of giant molecular
cloud during star formation and also in galaxy formation. 

The article is arranged in the following manner. In the next section
we have developed the basic mathematical formalism for a two-dimensional
hypothetical white dwarf star. In section-III, we have investigated
the gross properties of white dwarf stars in two-dimension. In
section-IV, the stability of two-dimensional white dwarfs with
ultra-relativistic electrons as one of the constituents 
have been studied. Finally in the last
section we have given the conclusion of this work.
\section{Basic Formalism}
In this section we start with the conventional form of Poisson's equation,
given by \cite{MAX} (see also \cite{SP})
\begin{equation}
\nabla^2\phi=2\pi G\rho
\end{equation}
where $\phi$ is the gravitational potential, $\rho$ is the surface density 
of matter and for the two-dimensional
scenario $\nabla^2$ has to be expressed in its two-dimensional form.  Let us
assume that the bulk object in two dimensional 
geometry is an hypothetical white dwarf star for which the inward
gravitational pressure  is balanced by the
outward degeneracy
pressure of the two dimensional electron gas. The mass of the object 
is coming from the heavy nuclei distributed in two-dimensional
bounded geometry.

Now the Fermi energy for a two dimensional electron gas is given by 
\begin{equation}
\mu_e=\frac{p_{F_e}^2}{2m_e}
\end{equation}

where $p_{F_e}$ and $m_e$ are respectively the Fermi momentum and the
mass of the electrons. Further, the number of electrons per unit surface 
area is given by
\begin{equation}
n_e=\frac{1}{2\pi\hbar^2}p_F^2
\end{equation}
Hence the electron Fermi energy is
\begin{equation}
\mu_e=\frac{\pi\hbar^2}{m_e}n_e
\end{equation}
Let $m_p$ is the baryon mass per electron \cite{LL}, then it can very easily be shown that 
\begin{equation}
\mu_e=\frac{\pi\hbar^2}{m_em_p}\rho
\end{equation}
where $\rho$ is the mass density (mass per unit area) of the matter.
Then following \cite{LL}, the Thomas-Fermi condition is given by
\begin{equation}
\mu_e+m_p\phi=~\rm {constant}
\end{equation}

Since $\phi$ is a function of radial coordinate $\vec r$, the chemical
potential $\mu_e$ and the matter density $\rho$ also depend on the radial 
coordinate $\vec r$.
Then replacing the gravitational potential $\phi$ with the electron Fermi energy $\mu_e$ from eqn.(6), the Poisson's
equation (eqn.(1)) can be written as
\begin{equation}
\nabla^2\mu_e=-2\pi Gm_p\rho
\end{equation}
Assuming circular symmetry i.e., independent of $\theta$ coordinate, 
replacing $\rho(r)$ by $\mu_e(r)$ from eqn.(5) and expressing $\nabla^2$ in
two-dimensional polar form, we have from eqn.(7) 
\begin{equation}
\frac{d^2\mu_e}{dr^2}+\frac{1}{r}\frac{d\mu_e}{dr}=-\frac{2Gm_em_p^2}{\hbar^2}\mu_e
\end{equation}
Substituting $x=r\lambda^{\frac{1}{2}}$, where $x$ may be called 
the scaled radial coordinate with the constant
\begin{equation}
\lambda=\frac{2Gm_em_p^2}{\hbar^2}
\end{equation}
we can write down the Poisson's equation in the following form
\begin{equation}
\frac{d^2\mu_e}{dx^2}+\frac{1}{x}\frac{d\mu_e}{dx}+\mu_e=0
\end{equation}
It is quite obvious that in this case the solution is given by \cite{AS}
\begin{equation}
\mu_e(x)=AJ_0(x)
\end{equation}
where $J_0(x)$ is the ordinary Bessel function of order zero. The
surface of the two-dimensional
bulk object is obtained from the first zero $x_s$(say) of the Bessel
function $J_0(x)$, i.e., 
\begin{equation}
\mu_e(x_s)=AJ_0(x_s)=0
\end{equation}
Hence the radius of the object is given by $R=x_s/\lambda^{\frac{1}{2}}$.
Further, at the centre, i.e., for $x=0$,
\begin{equation}
\mu_0(0)=AJ_0(0)=A=~~{\rm{constant}}
\end{equation}
Therefore the solution can also be written in the form
\begin{equation}
\mu_e(x)=\mu_e(0)J_0(x)
\end{equation}
Hence the central density is given by
\begin{equation}
\rho(0)=\frac{\lambda}{2\pi Gm_p}\mu_e(0)
\end{equation}
Therefore we can also write
\begin{equation}
\rho(x)=\rho(0)J_0(x)
\end{equation}
gives the variation of matter density with $r$
\section{Mass-Radius Relation}
With circular symmetry, the Poisson's equation in polar coordinate can also 
be written in the following form
\begin{equation}
\frac{1}{r}\frac{d}{dr}\left (r\frac{d\phi}{dr}\right )=2\pi G\rho
\end{equation}
Let us now integrate this differential equation with respect to r from $r=0$,
which is the centre to $r=R$, the surface of the bulk two dimensional 
stellar object. Then we have 
\begin{equation}
GM=R\frac{d\phi}{dr}\mid_{r=R}=x_s\frac{d\phi}{dx}\mid_{x=x_s}
\end{equation}
where 
\begin{equation}
M=\int_0^R 2\pi \rho rdr
\end{equation}
the mass of the object. Now expressing $\phi$ in terms of $\mu_e$ and using 
the solution for $\mu_e$ (eqn.(14)), we finally get 
\begin{equation}
GM=\frac{x_s\mu_e(0)}{m_p}J_1(x_s)
\end{equation}
where we have used the standard relation $dJ_0(x)/dx=-J_1(x)$ \cite{AS}. 
Therefore this transcendental equation (eqn.(20)) can be solved numerically to
get mass-radius relation for the two-dimensional stellar objects in
Thomas-Fermi model in the non-relativistic scenario. In fig.(1), we have shown 
graphically the form of mass radius relation for such objects. For the 
sake of illustration we have chosen the central density $\rho(0)$
in such a manner 
that the maximum mass of the object is $\sim 1.4M_{\odot}$.
The average density can also be obtained from the definition 
\begin{equation}
\overline\rho=\frac{M}{\pi R^2}
\end{equation}
Which can be expressed in terms of central density, and is given by
\begin{equation}
\overline\rho=\frac{2\rho(0)}{x_s}J_1(x_s)
\end{equation}
\section{Ultra-Relativistic Scenario}
In this section following \cite{LL} we have considered a two-dimensional hypothetical White dwarf star composed of
massive ions and degenerate ultra-relativistic electron gas in 
two-dimensional geometrical configuration. The stability of the object is governed by electron degeneracy pressure.
In the ultra-relativistic scenario the
energy of an electron is  given by the usual expression
\begin{equation}
\varepsilon=cp
\end{equation}
where $c$ is the velocity of light and $p$ is the electron momentum. 
The Fermi energy for this degenerate electron gas is then given by
\begin{equation}
\mu_e=cp_F
\end{equation}
with $P_F$, the electron Fermi momentum. Hence the  number density (the surface value) for ultra-relativistic electron 
gas can be written in the following form
\begin{equation}
n_e=\frac{p_F^2}{2\pi\hbar^2}
\end{equation}
Therefore the electron Fermi energy 
\begin{equation}
\mu_e=c\hbar(2\pi n_e)^{\frac{1}{2}}=c\hbar\left (\frac{2\pi\rho}{m_p}\right )^{\frac{1}{2}}
\end{equation}
where $n_e\approx\rho/m_p$ and $\rho$ is the mass density for such ultra-relativistic matter.

Then following eqn.(7), we have the Thomas-Fermi equation in two-dimension in 
the ultra-relativistic electron gas scenario
\begin{equation}
\frac{d^2\mu_e}{dx^2}+\frac{1}{x}\frac{d\mu_e}{dx}+\mu_e^2=0
\end{equation}
where $x$ is the scaled radial coordinate and the constant 
\begin{equation}
\lambda=\frac{Gm_p^2}{(\hbar c)^2}
\end{equation}
Unfortunately the non-linear differential equation given by eqn.(27) 
can not be solved analytically. To obtain its numerical solution we have used the standard four point Runge-Kutta
numerical technique and a code is written in FORTRAN 77 to solve eqn.(27) using 
the initial conditions (i) $\mu_e(0)=2\pi\hbar c\rho(0)/m_p$, at
$x=0$ (which indicates the origin of the two-dimensional white
dwarf), which is the maximum value of
$\mu_e$, then as a consequence 
(ii) $d\mu_e/dx=0$, the other initial condition. The surface of this bulk two-dimensional 
object is then obtained from the boundary condition $\mu_e(x_s)=0$,
with $x_s$ the scaled radius parameter. From the numerical solution of 
eqn.(27) with the initial conditions (i) and
(ii), the actual value of the radius can be obtained.
As a cross check, we have used MATHEMATICA in LINUX platform and got
almost same result.
In our formalism, instead of solving for the gravitational potential
in two-dimension, we have solved numerically for the electron
chemical potential with the initial conditions (i) and (ii). The
initial condition (i) depends on the central density $\rho(0)$, which
we have supplied by hand as input. In the non-relativistic case by
trial and error method we have fixed the value $\rho(0) \approx
10^9$g cm$^{-3}$ to get $M/M_\odot\sim 1.4$. Whereas in the
ultra-relativistic electron gas scenario, we have started from
$\rho(0)=10^6$g cm$^{-3}$ and gone upto $10^{12}$g cm${-3}$. Since
we are solving numerically for $\mu_e$ and the central density is
supplied by hand, the singularity at the origin ($x=0$) does not
appear in our analysis. Singularity at the origin also do not appear when
Lane-Emden equation (Newtonian picture) and TOV equation (general relativistic scenario) are solved numerically
by supplying the central density by hand.
In this case the mass-radius relation can also be derived from
eqn.(17) with $\phi$ replaced by $\mu_e$ from eqn.(6), the
Thoma-Fermi condition,  and is given by 
\begin{equation}
GM=-\frac{x_s}{m_p}\frac{d\mu_e}{dx}\mid_{x=x_s}
\end{equation}
Unlike the non-relativistic scenario, in this case the derivative
term at the surface has to be obtained from the numerical solution of Thomas-Fermi
equation (eqn.(27)). In fig.(2) we have shown the graphical form of 
mass radius relation for the ultra-relativistic
case. The qualitative nature of fig.(1) and fig.(2) are
completely different.
Unlike the non-relativistic situation, where the mass radius relation 
is obtained only for a particular
central density, in this case each point on the M-R curve corresponds a 
particular central density. This is obvious from the initial
condition (i), where $\rho(0)$ is the given central
density. Since the central
density has a fixed value in the non-relativistic scenario, the mass of the 
object increases with the increase of its
radius. To make these two figures more understandable, in fig.(3) and 
fig.(4) we have plotted the variations of mass and
radius respectively for the object against the central density in the ultra-relativistic scenario. In fig.(3) we have plotted 
the mass of the object expressed in terms
of solar mass, with the central density of the object expressed in terms of 
normal nuclear density. In fig.(4) we have shown the
variation of radius of the object with central density. From the
study of density profiles of the object in the ultra-relativistic
scenario, we have noticed that for the low value of the central
density, the numerical solution of eqn.(27) shows that the matter
density vanishes or becomes negligibly small  for quite large radius value, whereas for very
high central density, the matter density vanishes very quickly.
Therefore in this model the stellar objects with very low central
density are large in size and since the matter density is low enough,
the mass is also quite small. On the other hand for the large values of
central density, the objects are small in size, i.e., they are quite
compact in size but massive enough. This study explains the nature of
variations of mass and radius with central density for the
ultra-relativistic case. Further in ultra-relativistic scenario the average
density of matter inside the bulk two-dimensional stellar object is
given by
\begin{equation}
\overline\rho=\frac{\rho(0)}{x_s\mu_e^2(0)}\mid\frac{d\mu_e}{dx}\mid_{x=x_s}
\end{equation}
which exactly like the non-relativistic case also depends on the
central density and the radius of the object. However, unlike the
non-relativistic situation, here the average density depends on the  surface value of the gradient of electron Fermi energy, which has to
be obtained numerically from the solution of eqn.(27). 
\section{Conclusion}
In conclusion we would like to comment that although the formalism developed here is for a hypothetical stellar
object, which is basically an extension of standard three dimensional problem discussed in the book by Landau \&
Lifshitz, we strongly believe that it may be considered as interesting post-graduate level problem for the physics
students, including its numerical part. this problem can also be treated as a part of dissertation project for
post graduate physics students. The study of Thomas-Fermi model in
two-dimension for self-gravitating objects may be used for model
calculations of star formation and galaxy formation from giant
gaseous cloud. Finally, we believe that the problem solved here has some
academic interest. 

\noindent{\bf{Acknowledgment:}} We would like to thank the anonymous
referee for constructive criticism and pointing out some of the
usefull references.

\newpage
\begin{figure}
\psfig{figure=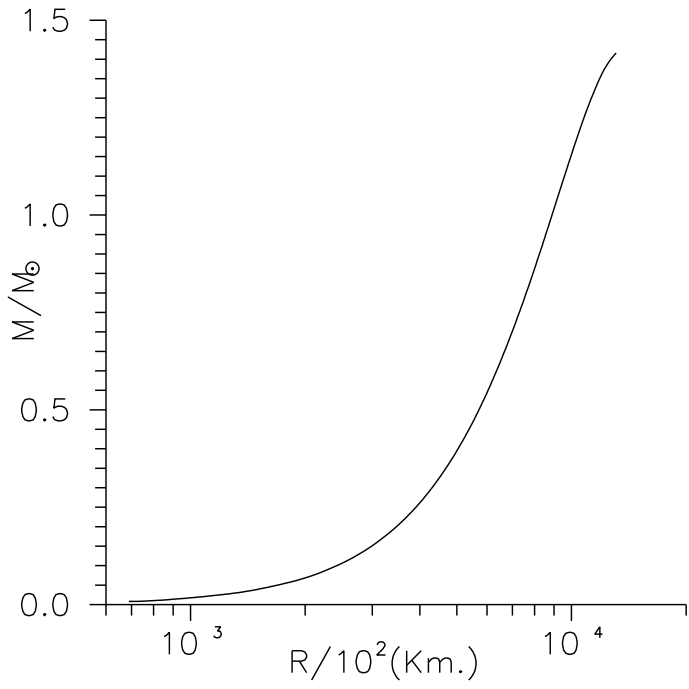,height=0.5\linewidth}
\caption{Mass-Radius relation for the non-relativistic case}
\end{figure}

\begin{figure}
\psfig{figure=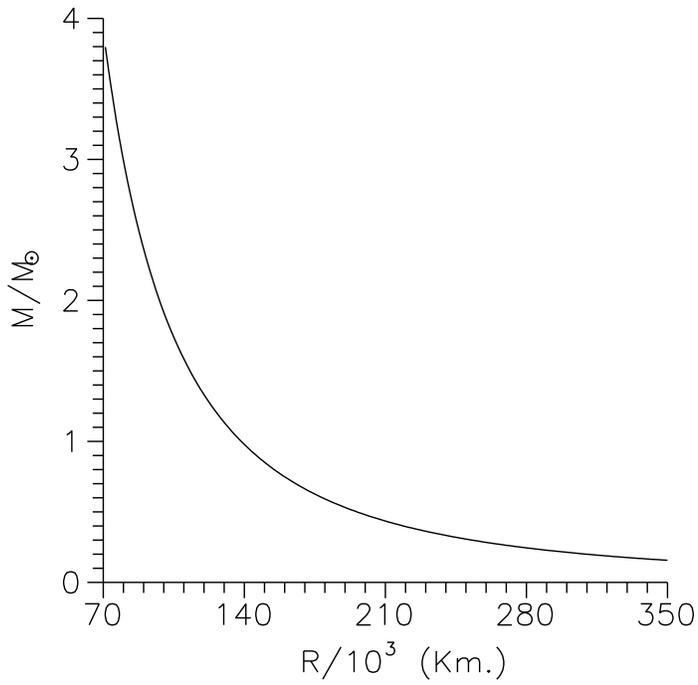,height=0.5\linewidth}
\caption{Mass-Radius relation for the ultra-relativistic case}
\end{figure}

\begin{figure}
\psfig{figure=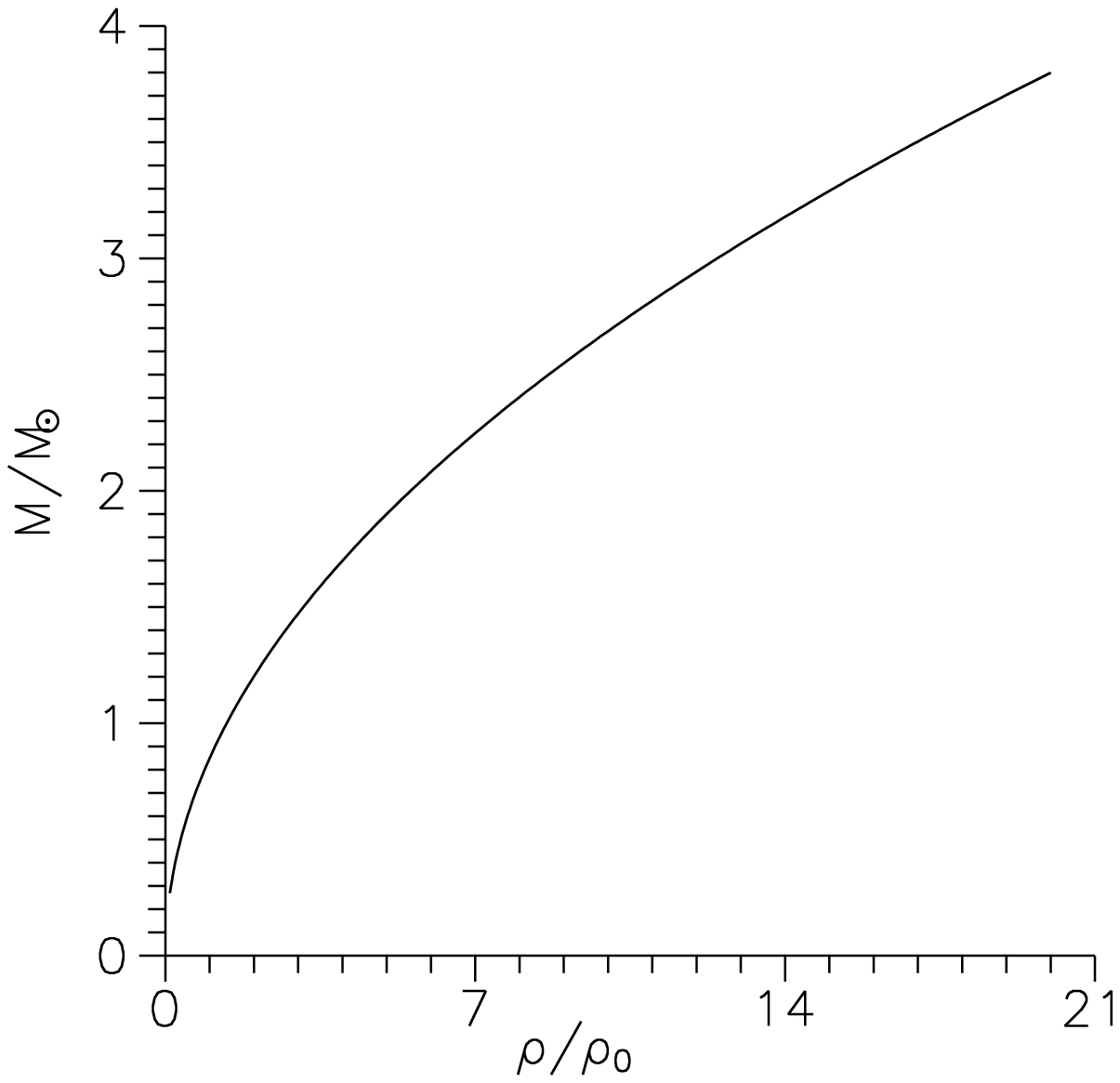,height=0.5\linewidth}
\caption{Variation of mass with density for the ultra-relativistic case}
\end{figure}

\begin{figure}
\psfig{figure=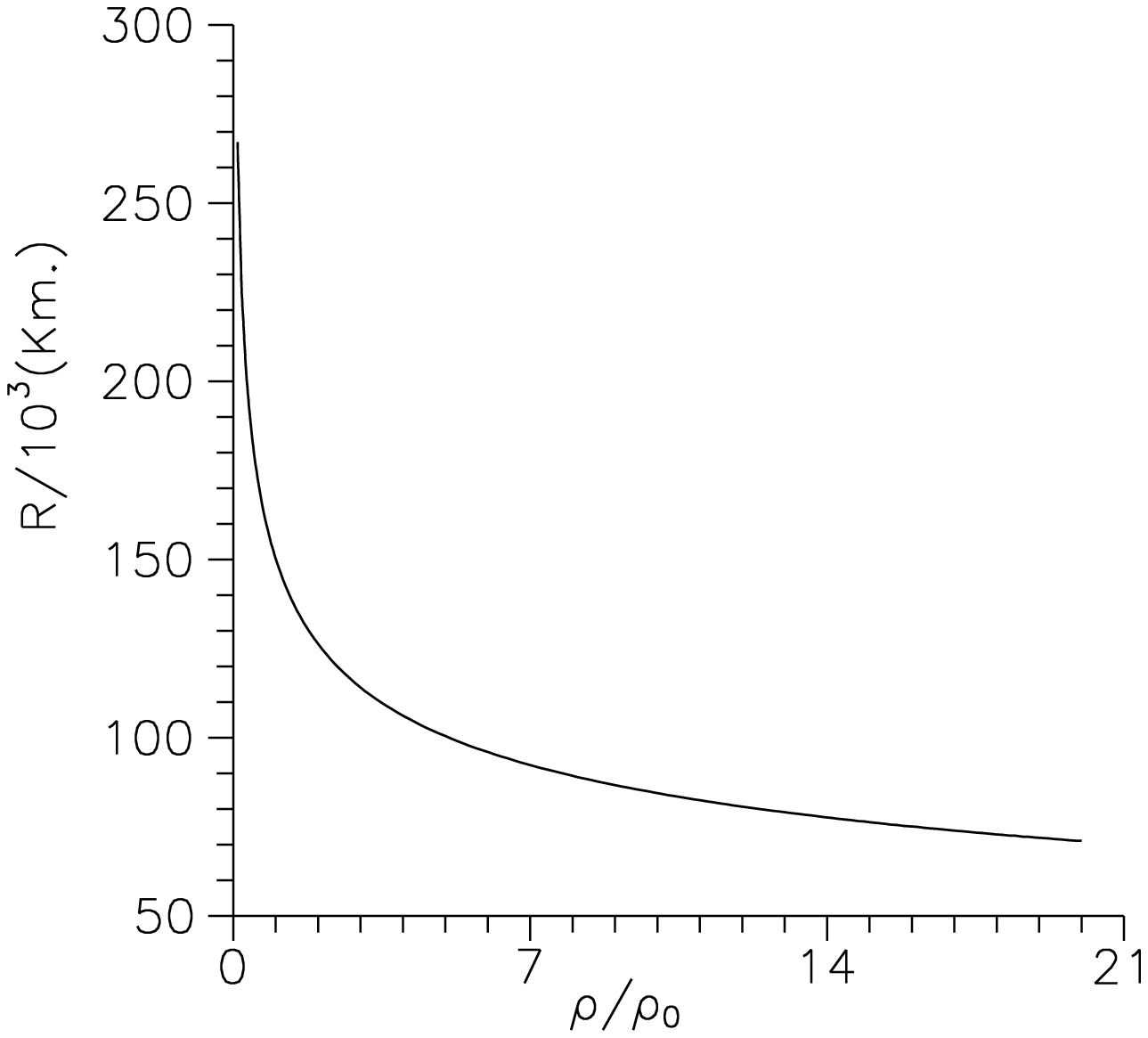,height=0.5\linewidth}
\caption{Variation of radius with density for the ultra-relativistic case}
\end{figure}
\end{document}